\documentclass[twocolumn,showpacs]{revtex4}

\usepackage{amsmath}
\usepackage{amssymb}

\begin{document}

\title{%
A generalization of Clausius inequality
for processes between nonequilibrium steady states
in chemical reaction systems}
\author{Tatsuo Shibata}
\affiliation{%
Research Institute for Mathematical Sciences,
Kyoto University,
Kyoto, 606-8502, JAPAN}
\date{submission:  December 21, 2000; print: \today}

\begin{abstract}
We investigate nonequilibrium chemical reaction systems
\null from the view point of steady state thermodynamics
proposed by Oono and Paniconi
[Prog.~Theor.~Phys.~Suppl. {\bf 130}, 29 (1998)].
The concentrations of some compounds are operated by an external system,
so that a transition \null from a steady state to other steady state takes place.
We show that an analogue of Clausius inequality holds macroscopically
for the operation processes.
This implies that
the second law of thermodynamics can be generalized,
including nonequilibrium steady states.
\end{abstract}


\maketitle


Most of the interesting and important chemical reactions
take place under nonequilibrium conditions,
for instance,
in biological systems
and in chemical plants.
When the concentration of a compound is changed by an external system,
how do the concentrations of the other compounds respond?
What kinds of restrictions are imposed on possible physical processes?
Beyond individual systems,
if there is a phenomenological framework for nonequilibrium states
to study such questions,
it is as valuable as equilibrium thermodynamics.

A lot of efforts have been done
to describe nonequilibrium systems
\null from thermodynamic viewpoints~\cite{Non}.
Oono and Paniconi have proposed
a phenomenological framework,
{\em steady state thermodynamics}~(SST)~\cite{Oono},
for nonequilibrium steady states
in a parallel form to equilibrium thermodynamics.
A significant characteristic of equilibrium thermodynamics
is the second law which asserts
the existence of {\em quasiequilibrium} and {\em non-quasiequilibrium} processes,
that is responsible for the evolution and stability criteria.
Similarly, they consider that
even for the processes between nonequilibrium steady states
``{\em quasisteady}'' and ``{\em non-quasisteady}'' processes should exist.
Since the system steadily generates dissipations,
we need to carefully extract a quantity
relevant to the operation to change the steady state.
For such a quantity,
they propose the ``{\em excess heat}'' absorption
$
Q_{ex}=Q-Q_{hk}
$
where $Q$ is the heat absorbed by the system,
and $Q_{hk}$ is the ``{\em house-keeping}'' heat absorption
to maintain the steady state.
In an appropriate ``{\em quasisteady}'' limit,
the quantity should be equal to
the difference of a pertinent ``{\em thermodynamic function}'' of SST.

Whereas ultimate justification of the framework
remains with its experimental analysis,
it may be worthwhile to explore
several models for the validity of SST.
Sekimoto and Oono studied a simple Langevin system
so that they found the house-keeping heat and a thermodynamic function
for the system~\cite{Sekimoto}.
Hatano found the minimum work principle
for nonequilibrium steady states assuming a particular
condition~\cite{Hatano}.
Then Hatano and Sasa
investigated a Brownian particle driven by an external force~%
\cite{HatanoSasa}.
For a process operated by an external system,
they found the generalized Jarzynski equality~\cite{Jarzinski},
$\langle\exp{\{\beta Q_{ex}+\phi(x_f;a_f)-\phi(x_i;a_i)\}}\rangle=1$
where $\phi(x;a)=\log{p(x;a)}$
with $p(x;a)$ the stationary probability of
the steady state given a parameter~$a$,
$a_i$ and $a_f$ are the parameters
of the initial and final state of the process,
$\beta$ is the inverse temperature,
and
$\langle\cdot\rangle$ indicates the ensemble average over all possible paths.
Then, identifying $S(a)=-\langle\phi(x,a)\rangle=-\int p(x;a)\log{p(x;a)}dx$
as the entropy function of SST,
they found generalized Clausius inequality
$
\beta Q_{ex}\leq S(a_f)-S(a_i).
$
They also proposed that the {\em house-keeping} heat
is related to the degree of breakdown of the detailed balance
condition.
%

In this Letter,
we investigate nonequilibrium chemical reaction systems
\null from the viewpoint of SST.
We consider ``ideal'' reaction systems, i.e.,
the mixture and reactions are homogeneous in space,
the non-reactive collisions of molecules occur frequently
between the successive reactions,
and
the time scale of the reactions
and that of the internal degrees of freedom of the molecules
are separated.
Then 
the system can be described by
Markov jump processes~\cite{vanKampen}.
We show that the similar argument developed in~\cite{HatanoSasa} 
also holds in such systems.
First we show that
a generalized Jarzynski equality holds for Markov jump processes.
Connecting the equality with an energetic interpretation,
we show a generalized Clausius inequality,
which is the manifestation of non-quasisteady processes.


Consider a vessel with its volume~$V$,
in which $(M+N)$~chemical compounds~%
${\mathsf{A}}_i~(i=1,\cdots,M)$
and
$\mathsf{X}_j~(j=1,\cdots,N)$ exist.
Among these chemical compounds,
the following chemical reactions take place:
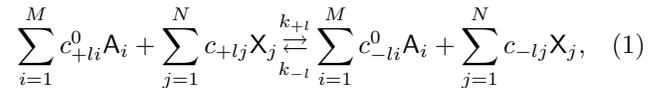
\begin{equation}
\sum_{i=1}^M
c_{+li}^{0}{\mathsf{A}}_i
+
\sum_{j=1}^N
c_{+lj}
\mathsf{X}_j
\underset{k_{-l}}{{\overset{k_{+l}}{\rightleftarrows}}}
\sum_{i=1}^M
c_{-li}^{0}{\mathsf{A}}_i
+
\sum_{j=1}^N
c_{-lj}\mathsf{X}_j,
\end{equation}
with $l=1,\cdots,L$.
Let us call the forward reactions reaction~$+l$
while the backward reactions reaction~$-l$.
We assume that both the rate constants~%
$k_{+l}$ and $k_{-l}$
of the reaction~$+l$ and reaction~$-l$, respectively,
are positive.
The stoichiometric coefficients~%
$c_{\pm li}^0$ and  $c_{\pm lj}$
are non-negative integer.
The concentrations of ${\mathsf{A}}_i$'s are kept~$a_i$.
The number of molecules~$\mathsf{X}_j$ is denoted by~$n_j$.
The state of the system is denoted by $n=(n_1,\cdots,n_N)$.
Let $\nu_{li}^0$ be $\nu_{+lj}^0=-\nu_{-lj}^0=c_{-lj}^0-c_{+lj}^0$
and $\nu_{lj}$ be $\nu_{+lj}=-\nu_{-lj}=c_{-lj}-c_{+lj}$.
When the reaction~$l\in[-L,\cdots,-1,1,\cdots,L]$ takes place,
the state jumps \null from $n$ to $n+\nu_l=(n_1+\nu_{l1},\cdots,n_N+\nu_{lN})$.

For~$\mathsf{A}_i$'s,
one can consider that
these compounds are exchanged
between the vessel and its surroundings,
and the exchange process is sufficiently fast
so that the concentrations are well controlled.
Although we neglect to consider the exchange process for simplicity,
it is always possible to take the process into account.
We then consider the situation
where an external system
operates the system in time
by changing some of the concentrations~$a_i$'s
during the times~$[0,t]$.
The set of such time-dependent concentrations at time~$s$
are denoted by $a(s)$,
and $a(s)$ is changed according to
a given operation protocol~$a(s)_{0\leq s\leq t}$.

The probability~$p(n,s)$ that the system is in the state~$n$
at time $s$ follows the master equation,~
\begin{equation}
\frac{\partial p(n,s)}{\partial s}
=
\sum_{l=-L}^{L}
\left(\mathbb{E}^{\nu_{l}}\!-\!1\right)
W_{l}(n,n+\nu_{l};a(s))
p(n,s)
\label{eq:master}
\end{equation}
where
$\mathbb{E}^{\nu_l}$ is an operator
defined by its effect on an arbitrary function $f(n)$,
$
\mathbb{E}^{\nu_l}f(n)=f(n_1-\nu_{l1},\cdots,n_N-\nu_{lN}),
$
and
$W_{l}(n,n+\nu_{l};a)$
is the transition probability, given $a$,
that the reaction~$l$ takes place per unit time~\cite{vanKampen},
\begin{equation}
W_{l}(n,n+\nu_{l};a)
=
V k_{l}
\prod_{i=1}^M
a_i^{c_{li}^0}
\prod_{j=1}^N
\biggl(
\frac{n_j!/(n_j-c_{lj})!}{V^{c_{lj}}}\biggr)
\label{eq:transition}
\end{equation}
with $l\in[-L,\cdots,-1,1,\cdots,L]$.
We suppose that
if the time-dependent parameters are virtually fixed at $a$,
the probability distribution function approaches
a unique stationary probability distribution function~$p_s(n;a)$
that satisfies~%
$
\sum_{l=-L}^{L}
\left(\mathbb{E}^{\nu_{l}}\!-\!1\right)
W_{l}(n,n+\nu_{l};a)
p_s(n;a)=0
$.

Let us consider a particular state at time~$s$ denoted by~$n(s)$,
and a particular path denoted by~$n(s)_{0\leq s\leq t}$.
Starting \null from an initial state~$n(0)$,
$n(s)$ is given by
\begin{equation}
n_j(s)=n_j(0)+\int_0^s\sum_{l=-L}^L\nu_{lj}\xi_{l}(u)du
\label{eq:number}
\end{equation}
where each $\xi_{l}(s)$ is a sequence of $\delta$-function,
located at those times when the reaction~$l$ takes place.

For these processes,
we consider an energetic interpretation.
In the present model system,
we simply suppose that
a molecule of each compound has
the energy~$E_i^0$ of $\mathsf{A}_i$ or~$E_j$ of $\mathsf{X}_j$
and no interaction energy among molecules is taken into account.
Under equilibrium conditions,
the states of the system should obey the grand canonical ensemble,
and the detailed balance condition should hold.
This leads to
a relation between rates of a transition and its reverse, as
\begin{equation}
\frac{k_{+l}}
{k_{-l}}
=
\prod_{i=1}^M
\left(v_0e^{\beta E_i^0}\right)^{-\nu_{li}^0}
\prod_{j=1}^N
\left(v_0e^{\beta E_j}\right)^{-\nu_{lj}}
\label{eq:RateConstant}
\end{equation}
where $v_0$ is a certain constant of the dimension of~$V$.

For a particular state~$n$,
the internal energy of the system~$\mathcal{E}$ is given by~%
$\mathcal{E}=\sum_{j=1}^{N}E_jn_j$.
The internal energy change~$\Delta\mathcal{E}(t)$ is given by
\begin{equation}
\Delta\mathcal{E}(t)
=\mathcal{E}(t)-\mathcal{E}(0)
=
\sum_{j=1}^{N}
\int_0^t
\sum_{l=-L}^L
E_j\nu_{lj}\xi_{l}(u){du}.
\end{equation}
Here the energy of~$\mathsf{A}_i$'s
is not included in~$\mathcal{E}$
for the following reason.
Since in the present consideration
the concentrations of~$\mathsf{A}_i$'s
completely follow the parameters~$a_i$'s at every moment,
the role of~$\mathsf{A}_i$'s is nothing more than that of reservoirs.
If one would consider this unphysical,
the injection and extraction processes should be took into account.


Next we show that the generalized Jarzynski equality,
which shall be given in Eq.(\ref{eq:IdentityPhi}),
holds in the Markov jump processes
generated according to~$W_l$.
%
%
The idea to prove the equality has been developed first by
Crooks under equilibrium conditions~\cite{Crooks}.
Then, Hatano and Sasa generalize the argument
for Langevin systems under nonequilibrium conditions~\cite{HatanoSasa}.

Consider that a particular path~$n(s)_{0\leq s\leq t}$,
which is a sequence of states
\begin{equation}
n(0)
\stackrel{l_1}{\rightarrow}
n(t_1)
\stackrel{l_2}{\rightarrow}
\cdots
\stackrel{l_n}{\rightarrow}
n(t_n),
\label{eq:reaction}
\end{equation}
with $t_i$ a jump time when
the system jumps \null from $n(t_{i-1})$ to $n(t_i)=n(t_{i-1})+\nu_{l_i},~%
(i=1,\cdots,n$, $t_n\leq t<t_{n+1})$.

If the system is in the state~$n$ at time~$s$,
the probability that
a transition takes place per unit time is given by
$W(n;a(s))=\sum_{l=-L}^LW_{l}(n,n+\nu_l;a(s))$.
The system stays in~$n$ until $s'=s+\tau$,
where $\tau$ is an independent random variable
distributed according to the density function~%
$W(n;a(s+\tau))\exp{\left\{-\int_{s}^{s+\tau}W(n;a(u))du\right\}}$.
Then, at $s'$
the system jumps to $n+\nu_l$ according to the probability~%
$W_l(n,n+\nu_l;a(s'))/W(n;a(s'))$.

Suppose that at time~$0$
the system is in a steady state.
Then,
the probability~%
$\mathcal{P}[t]=\mathcal{P}[n(s)_{0\leq s\leq t};a(s)_{0\leq s\leq t}]$
that we have the path~$n(s)_{0\leq s\leq t}$
is given by
\begin{eqnarray}\nonumber
\mathcal{P}[t]
&=&
p_s(n(0);a(0))
\prod_{i=1}^n
W_{l_i}(n(t_{i-1}),n(t_i);a(t_{i}))
\\
&&\times
e^{-\int_{0}^{t}W(n(u);a(u))du},
\end{eqnarray}
where $t_0=0$.

Consider the transition probability~$\widetilde{W}_l(n,n+\nu_l;a)$,
defined by
\begin{equation}
\widetilde{W}_l(n,n+\nu_l;a)
=
\frac{p_s(n+\nu_l;a)}{p_s(n;a)}
W_{-l}(n+\nu_l,n;a).
\label{eq:dualW}
\end{equation}
%
%
If the detailed balance condition holds,
$\widetilde{W}_l=W_l$.
For a given path~$n(s)_{0\leq s\leq t}$ generated by~${W}_l$,
consider the path~$\widetilde{n}(s)_{0\leq s\leq t}$ generated by~$\widetilde{W}_l$
given by~$\widetilde{n}(s)=n(t-s)$
with a parameter protocol~$\widetilde{a}(s)=a(t-s)$.
The jump times for~$\widetilde{n}(s)_{0\leq s\leq t}$ is given by~%
$\tilde{t}_i=t-t_{n-i+1}$
at which 
the transition takes place
from $\widetilde{n}(\tilde{t}_{i-1})$
to $\widetilde{n}(\tilde{t}_i)
=\widetilde{n}(\tilde{t}_{i-1})+\nu_{\tilde{l}_i}$~%
with~$\tilde{l}_i=-l_{n-i+1}$.
The probability~%
$\widetilde{\mathcal{P}}[t]
=
\widetilde{\mathcal{P}}(\widetilde{n}(s)_{0\leq s\leq t};
\widetilde{a}(t)_{0\leq s\leq t})$
that we have the path~$\widetilde{n}(s)_{0\leq s\leq t}$
is also calculated as is the case of~${\mathcal{P}}[t]$.
Here, we suppose that $\widetilde{n}(s)$ also starts from a steady state.
Now we compare $\widetilde{\mathcal{P}}[t]$ with ${\mathcal{P}}[t]$.
Let us introduce the quantity~$\phi(n;a)$,
\begin{equation}
p_s(n;a)=\exp{\phi(n;a)}.
\label{eq:phi}
\end{equation}
Using Eq.(\ref{eq:dualW}),
and noting
$
\widetilde{W}(n;a)=
\sum_{l=-L}^L\widetilde{W}_{l}(n,n+\nu_l;a)
=W(n;a),
$
$\widetilde{\mathcal{P}}[t]$
has a density
relative to 
$\mathcal{P}[t]$
as
\begin{equation}
\widetilde{\mathcal{P}}[t]
=
e^{%
\phi(n(t),a(t))-\phi(n(0),a(0))
+\Phi[t]}
\mathcal{P}[t],
\label{eq:PtildeP}
\end{equation}
where
$\Phi[t]=\Phi[n(s)_{0\leq s\leq t};a(s)_{0\leq s\leq t}]$ is the quantity
\begin{equation}
\Phi[t]
=
\int_0^t
\biggl\{
\phi(n(s_{-0});a(s))-\phi(n(s);a(s))
\biggr\}\xi(s)ds
\label{eq:defPhi}
\end{equation}
with $\xi(s)=\sum_{l=-L}^L\xi_{l}(s)$
and  $n(s_{-0})$ the state just before the jump to $n(s)$ at time~$s$,
i.e.,
the state after the previous jump event.
By summing Eq.(\ref{eq:PtildeP}) over all possible paths,
we obtain the generalized Jarzynski equality
\begin{equation}
\bigl\langle
\exp{\left\{
\phi(n(t),a(t))-\phi(n(0),a(0))
+\Phi[t]
\right\}
}
\bigr\rangle
=1
\label{eq:IdentityPhi}
\end{equation}
with Eqs.(\ref{eq:phi}) and (\ref{eq:defPhi}),
where
$\langle\cdot\rangle$ indicates the ensemble average
over all possible paths.

It may be worth mentioning that
if one consider that
the path~$\widetilde{n}(s)_{0\leq s\leq t}$
is generated according to~$W_l$ instead of
$\widetilde{W}_l$,
the fluctuation theorem is obtained by comparing the path
probabilities~\cite{FluctuationTheorem,HatanoSasa}.

Eq.(\ref{eq:IdentityPhi})  leads to the relation
\begin{equation}
\left\langle\Phi[t]\right\rangle
\leq
-\langle\phi(t)\rangle-(-\langle\phi(0)\rangle),
\label{eq:AfterJenzen}
\end{equation}
where $-\langle\phi(s)\rangle=-\sum_np_s(n;a(s))\log{\{p_s(n;a(s))\}}$.
When the change of $a(s)$ is so slow
that $p(n,s)$ can be regarded as $p_s(n;a(s))$,
the equality holds in Eq.(\ref{eq:AfterJenzen}).

Next we interpret the inequality~(\ref{eq:AfterJenzen})
\null from energetic point of view
and show the existence of  non-quasisteady processes.
Let $\zeta_l(n,n+\nu_l;a)$ be the quantity
that characterize
the degree of the breakdown
of the detailed balance condition,
given by
\begin{equation}
\frac{p_s(n,a)W_{l}(n,n+\nu_l;a)}
{p_s(n+\nu_l,a)W_{-l}(n+\nu_l,n;a)}
=
e^{-\zeta_l(n,n+\nu_l;a)}.
\label{eq:breakdetailedbalance}
\end{equation}
Obviously, if the detailed balance condition holds,
$\zeta_l(n,n+\nu_l;a)=0$.
Substituting Eqs.(\ref{eq:transition}), (\ref{eq:RateConstant})
and (\ref{eq:breakdetailedbalance})
into Eq.(\ref{eq:defPhi}),
$\Phi[t]$ can be rewritten by
\begin{eqnarray}\nonumber
\Phi[t]
=
{\beta}\Delta\mathcal{E}(t)
+
{\beta}\int_0^t\sum_{l=-L}^L\sum_{i=1}^M\nu_{li}^0\mu_i^0(s)\xi_{l}(s)ds
\\\nonumber
-
\int_0^t\sum_{l=-L}^L
\zeta_{l}(n(s)-\nu_l,n(s);a(s))\xi_l(s)
ds
\\
+
\int_0^t\sum_{l=-L}^L
\log{\biggl(\prod_{j=1}^N
\frac{v_0^{\nu_{lj}}}{V^{\nu_{lj}}}
\frac{n_j(s)!}{(n_j(s)-\nu_{lj})!}
\biggr)}\xi_{l}(s)ds
\label{eq:Phi}
\end{eqnarray}
with
$
\mu_i^0(s)=E_i^0+\frac{1}{\beta}\log{\left\{v_0a_i(s)\right\}}
$,
the chemical potential of~$\mathsf{A}_i$.
%
%
For the first three terms  on the r.h.s.,
we introduce the quantity 
$\mathcal{D}_{ex}[t]=\mathcal{D}_{ex}[n(s)_{0\leq t};a(s)_{0\leq t}]$
given by
\begin{equation}
\!\mathcal{D}_{ex}[t]
\!=\!
\Delta\mathcal{E}(t)
\!+\!
\!\int_0^t\!\!\sum_{l=-L}^L\!\sum_{i=1}^M\nu_{li}^0\mu_i^0(s)\xi_{l}(s)ds
\!-\!
\mathcal{D}_{hk}[t].
\label{eq:Dex}
\end{equation}
where
$\mathcal{D}_{hk}[t]
=
\mathcal{D}_{hk}[n(s)_{0\leq s\leq t};a(s)_{0\leq s\leq t}]$
is define by
\begin{equation}
\!\!\!
\mathcal{D}_{hk}[t]
=
\frac{1}{\beta}
\int_0^t\sum_{l=-L}^L
\zeta_{l}(n(s)-\nu_l,n(s);a(s))\xi_l(s)ds.
\label{eq:house-keeping}
\end{equation}
Note that the fourth term on the r.h.s. of Eq.(\ref{eq:Phi})
is rewritten by
$
\log{\bigl\{%
\prod_{j=1}^N%
\left(%
\frac{v_0}{V}%
\right)^{n_j(s)}%
{n_j(s)!}%
\bigr\}}\bigr|_{s=0}^{s=t}.
$
Substituting Eq.(\ref{eq:Phi}) into Eq.(\ref{eq:AfterJenzen}),
with Eqs.(\ref{eq:Dex}) and (\ref{eq:house-keeping}),
we obtain the main result of this Letter
\begin{equation}
\beta\Bigl\langle\mathcal{D}_{ex}[t]\Bigr\rangle
\leq
\mathcal{S}(a(t))-\mathcal{S}(a(0)),
\label{eq:Clausius}
\end{equation}
where $\mathcal{S}(a)$ is the quantity define by
\begin{equation}
\!\!\!\!
\mathcal{S}(a)
=
-\sum_np_s(n;a)
\log{\biggl\{
\prod_i
\left(\frac{v_0}{V}\right)^{n_i}\!\!\!{n_i!}
p_s(n;a)
\biggr\}}.
\label{eq:entropy}
\end{equation}
In order to make
$\mathcal{D}_{ex}[t]$ in (\ref{eq:Dex}) transparent,
we introduce $\mathcal{A}_l$ the chemical affinity of the
reaction~$l$ as
\begin{equation}
\mathcal{A}_l=
-\sum_{i=1}^M\nu_{li}^0\mu_i^0
-\sum_{j=1}^N\nu_{li}\mu_i
=-\mathcal{A}_{-l}
\label{eq:affinity}
\end{equation}
with
$\mu_j(s)$ the chemical potential of~$\mathsf{X}_j$.
When $\mathcal{A}_l>0$, the reaction~$l$ proceeds on average. 
Under equilibrium conditions,
$\mathcal{A}_l=0$.
Using Eq.(\ref{eq:number}),
$\mathcal{D}_{ex}[t]$ is rewritten by
\begin{eqnarray}\nonumber
\!\mathcal{D}_{ex}[t]
&=&
\Delta\mathcal{E}(t)
\!-\!
\!\int_0^t\!\!\sum_{j=1}^N\!\mu_jdn_j(s)\\
&&
\!-\int_0^t\!\sum_{l=-L}^L\mathcal{A}_l\xi_l(s)ds
-\mathcal{D}_{hk}[t].
\label{eq:DexII}
\end{eqnarray}

Let us first see the consistency
between~(\ref{eq:Clausius}) and equilibrium thermodynamics.
Under equilibrium conditions
$\mathcal{D}_{hk}[t]=0$
and
$\mathcal{A}_l=0$.
Thus,
the first two terms on the r.h.s of~(\ref{eq:DexII}) 
are relevant to equilibrium thermodynamics.
By considering~$\mathcal{S}(a)$ as
the {\em equilibrium thermodynamic entropy},
(\ref{eq:Clausius}) is nothing but
Clausius inequality.

Under nonequilibrium conditions,
the third term on the r.h.s. of~(\ref{eq:DexII})
includes the continuous dissipation,
even when system's parameters are kept constants.
In order to discuss
the thermodynamics with respect to operations upon the system,
we need to carefully extract the quantities
relevant to the change in the macroscopic state.
When $da/dt$ is so small
that $p(n,s)$ can be regarded as $p_s(n;a(s))$,
$\langle\mathcal{D}_{ex}[t]\rangle$
coincides with the difference of~$\mathcal{S}(a)$
between the initial and final state of a process.
Let us then consider
$\langle\mathcal{D}_{ex}[t]\rangle$
as the quantity relevant to the operations,
and $\mathcal{S}(a)$ as the thermodynamic function
that characterize the nonequilibrium steady states.
The {\em quasisteady process} is realized
in the limit~$1/t\rightarrow0$,
given the parameter protocol~$a(s)=\hat{a}(s/t)$
with a given parameter path~$\hat{a}(\tau)_{0\leq\tau\leq1}$.
When $1/t$ is finite,
$\langle\mathcal{D}_{ex}[t]\rangle$ is smaller than
the difference of~$\mathcal{S}(a)$
and
the processes are considered to be {\em non-quasisteady processes}.
In~(\ref{eq:Dex}),
$-\langle\mathcal{D}_{hk}\rangle$
is regarded as the dissipation called the {\em house-keeping dissipation}
that keeps the system far \null from equilibrium~\cite{HatanoSasa}.
In this way,
we find a similar structure
in nonequilibrium steady states
to equilibrium thermodynamics,
and consider Eq.(\ref{eq:Clausius})
as a generalization of Clausius inequality.

In the definition of
the house-keeping dissipation~(\ref{eq:house-keeping})
and hence
the steady state entropy~(\ref{eq:entropy}) 
there is some arbitrariness.
So far no systematic discussion exists
in order to justify these definitions under nonequilibrium conditions.
However,
in some cases
we identify
$
-\left\langle
\mathcal{D}_{hk}[n(s)_{0\leq t}]
\right\rangle
$
as so-called entropy production.
When $p_s(n;a)$ is a Poisson distribution
or
for each compound
a grand canonical distribution can be supposed~(local equilibrium),
$\mathcal{D}_{hk}$ is written by
\begin{equation}
-
\left\langle
\mathcal{D}_{hk}[n(s)_{0\leq t}]
\right\rangle
\simeq
\left\langle
\int_0^t\sum_{l=-L}^L\mathcal{A}_l\xi_l(s)ds
\right\rangle
\label{eq:entropyproduction}
\end{equation}
where $\mu_j(s)$ in $\mathcal{A}_l$ is defined by
$\mu_j(s)=E_j+\frac{1}{\beta}
\log{\left\{v_0\langle n_j(a(s))\rangle/V\right\}}$
with $\langle n_j(a)\rangle=\sum_nn_jp_s(n;a)$.
The r.h.s.
of Eq.(\ref{eq:entropyproduction})
is identified as the entropy production~\cite{Non}.
In such cases,
although the condition is nonequilibrium,
$\mathcal{D}_{ex}$ can be reduced again as
\begin{equation}
\mathcal{D}_{ex}[t]=
\Delta\mathcal{E}(t)
-
\int_0^t\sum_{j=1}^N\mu_jdn_j(s),
\end{equation}
so that the same form as in equilibrium thermodynamics
is obtained.
This implies that
at least near equilibrium
the same relation with equilibrium thermodynamics
can be applicable to nonequilibrium system
by extending the definition of entropy.

If we compare the inequality~(\ref{eq:Clausius})
with
the previous result in~\cite{HatanoSasa},
$\beta Q_{ex}\leq S(a_f)-S(a_i)$
with~$Q_{ex}=Q-Q_{hk}$,
one might identify the first three terms on the r.h.s in~(\ref{eq:DexII})
as the heat absorbed by the system.
However, this is not the case,
since
in the present case
the dissipation without heat generation can take place
such as the mixing without energy conversion~\cite{ShibataSekimoto}.

We note that if each molecule is described individually,
$\mathcal{S}$ is rewritten by
$\mathcal{S}=-\sum_{\{n\}}p_s(\{n\})\log{p_s(\{n\})}$,
where $p_s(\{n\})$ is the probability
of the state~$\{n\}$ where the molecules are distinguished.

In the present Letter,
nonequilibrium chemical reactions have been studied
\null from the view point of steady state thermodynamics~\cite{Oono}.
First we have shown that the generalized Jarzynski equality
holds in the Markov jump process, and
then we have a generalization of Clausius inequality~(\ref{eq:Clausius}).
The argument presented here can be applicable to other systems
such as lattice gas systems.
Our choice of the connection
between the energetics and the kinetic parameters
is quite simple.
When the connection between them is much more complicated,
how our result will be modified remains to be clarified.
The experimental verification of the present result
is another significant problem.
Especially,
how the entropy function and the house-keeping dissipation can be measured
remains as a future problem.

The author thanks S.~Sasa, K.~Sekimoto and Y.~Takahashi for many discussions on
several related subjects,
T.~Hatano and S.~Sasa
for communicating their result before publication,
and S.~Takesue and M.~Matsuo
for useful discussions and comments on this manuscript.
He is grateful to Y. Itofuji for her continuous encouragements.

\end{document}